\documentclass[letterpaper]{article} 
\usepackage{aaai24}  
\usepackage{times}  
\usepackage{helvet}  
\usepackage{courier}  
\usepackage[hyphens]{url}  
\usepackage{graphicx} 
\usepackage{natbib}  
\usepackage{caption} 
\usepackage{algorithm}
\usepackage{algorithmic}

\usepackage{newfloat}
\usepackage{listings}
\DeclareCaptionStyle{ruled}{labelfont=normalfont,labelsep=colon,strut=off} 
\floatstyle{ruled}
\newfloat{listing}{tb}{lst}{}
\floatname{listing}{Listing}
\title{DQSSA: A Quantum-Inspired Solution for Maximizing Influence in Online Social Networks (Student Abstract)}
\author{
    Aryaman Rao\textsuperscript{\rm 1},
    Parth Singh\textsuperscript{\rm 1},
    Dinesh Kumar Vishwakarma\textsuperscript{\rm 1},
    Mukesh Prasad\textsuperscript{\rm 2}
}
\affiliations{
    \textsuperscript{\rm 1}Delhi Technological University, New Delhi, Delhi, India\\
    \textsuperscript{\rm 2}School of Computer Science, University of Technology Sydney, Ultimo, NSW 2007, Australia\\
    \url{aryaman26601@gmail.com}, \url{parth55singh@gmail.com}\\

}

\usepackage{bibentry}

\begin{document}

\maketitle
\begin{abstract}

Influence Maximization is the task of selecting optimal nodes maximising the influence spread in social networks. This study proposes a Discretized Quantum-based Salp Swarm Algorithm (DQSSA) for optimizing influence diffusion in social networks. By discretizing meta-heuristic algorithms and infusing them with quantum-inspired enhancements, we address issues like premature convergence and low efficacy. The proposed method, guided by quantum principles, offers a promising solution for Influence Maximisation. Experiments on four real-world datasets reveal DQSSA's superior performance as compared to established cutting-edge algorithms. 

\end{abstract}

\section{Introduction}
The Influence maximization (IM) problem revolves around extracting a subset of nodes, also known as seed nodes, from a social network that could maximize the influence spread \citep{chen2009im}. Practical applications of IM range from epidemiology to marketing, making it a popular subject of research due to its diverse real-world uses. Previous research on IM, although having produced significant advancements in effective seed node selection, suffers from excessive time complexity or inefficient influence spread. To overcome this, our approach strikes a balance between the network's influence spread and execution time, that is necessary to develop an efficient IM algorithm. Additionally, we tackle the problem of solutions often converging prematurely and getting trapped in local optimal points by incorporating quantum-inspired meta-heuristics (MH), drawing on quantum physics principles such as the Schrödinger Wave equation and Delta Potential fields \citep{ross2019review}. These approaches yield promising results, as elaborated in subsequent sections. The IM algorithm follows a sequential process, starting by identifying significant community structures using the Louvain algorithm. The selection of seed nodes, within communities, is facilitated through an objective function called the Local Influence Estimator or LIE. Once the solutions are obtained from the LIE function through DQSSA and other baseline approaches, an Independent Cascade (IC) diffusion model, depending on an infection probability hyper-parameter, is utilized to quantify the influence spread across a graph. 


\begin{figure*}
    \centering
    \includegraphics[scale = 0.35]{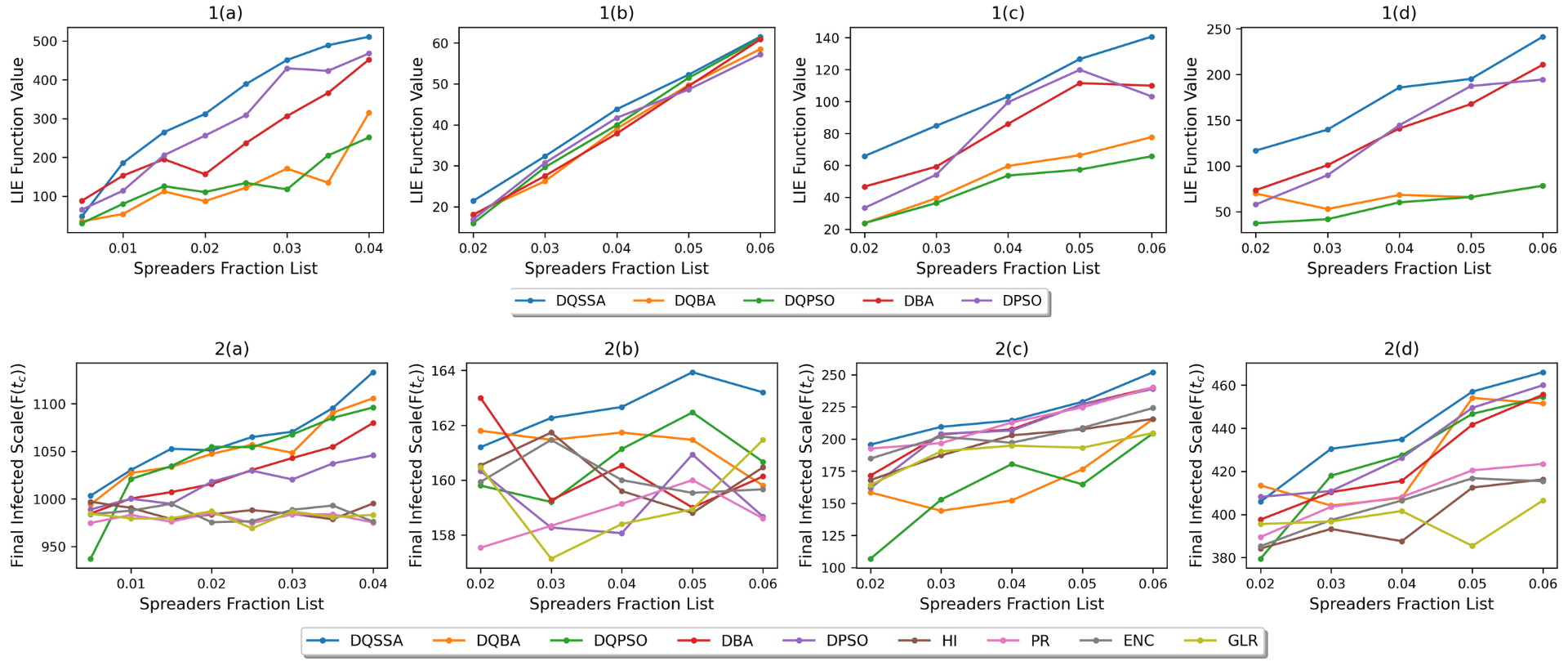}
    \caption{Experimental Analysis for DQSSA: (1) Value of LIE Function vs Spreader Fractions, for four meta-heuristic baseline algorithms. (2) Final Infected Scale (FIS), obtained from IC model simulation with infection probability 0.1. Plotted against four real-world data sets. These datasets include a) soc-hamsterster, b) jazz c) soc-wiki-Vote and d) email-univ.}
    \label{QSSA}
\end{figure*}

\section{Proposed Methodology}
\textbf{Community Detection}: Community structures are an integral part of information propagation in online social networks \citep{girvan2002community}. The nodes in these communities share similar behaviour and, are densely connected to the neighbours. The potential nodes in the selected communities undergo a budgeting process, ranking the nodes in the order of decreasing budgeting values where higher budget value indicates greater potential to spread information.\vspace{5pt}

\textbf{Quantum-based Salp Swarm Algorithm (Q-SSA)}: The Salp Swarm Algorithm (SSA), inspired by the swarms of salp organisms, faces challenges in extracting the global optimum due to premature convergence. To overcome this, we employed the QSSA approach, combining quantum-inspired techniques with SSA. In the transition to quantum-based MH algorithms, two essential strategies include Reverse Learning (RL) and Quantum Mutation (QM). RL(\ref{1}) is a cost-effective approach for solving high-dimensional problems combined with an effective internal search strategy to enhance performance. It further depends on two techniques: Opposite Point and Elite Opposite Solution, as explicated in \citep{chen2019qssa}. \vspace{3pt}
\begin{equation} \label{1} 
    X_{j}^{i}(t) = \mu \times(ub_j - lb_j) - x_j^i(t)
\vspace{5pt}
\end{equation}
In eq.(\ref{1}), $ub$ and $lb$ denote upper and lower bound and $\mu$ is a random number. Additionally, QM involves subjecting random solutions to a specific mutation criterion, with iterations continuing until specific conditions are met. \cite{@SAHA2015} elucidated the importance of mutation manipulation for global convergence using QM. In the eq.(\ref{2},\ref{3}), $\sigma$ is a constant, $x^i_j$ is a solution at any time $t$ and $X^i_j$ is the updated solution. Combining these diverse solutions randomly, increases the solution space diversity, facilitating the exploration of greater solutions for MH algorithms. 
\begin{equation} \label{2}
    X_j^i(t) = x_j^i(t) + \sigma \times \left( ub_j - x_j^i(t) \right)
\end{equation}
\begin{equation} \label{3}
    X_j^i(t) = x_j^i(t) + \sigma \times \left( x_j^i(t) - lb_j \right)
\vspace{5pt}
\end{equation}
\textbf{Discretization}: This process involves the conversion of traditional MH algorithms into their discrete forms \citep{zareie2020identification, rao2023influence}. The solution space of the IM is a set of discrete nodes, each with a specific capacity to propagate information effectively, best classified according to the LIE function. For identifying the optimal sets of nodes, continuous MH algorithms are not applicable. To address this, these algorithms are converted into their quantum-based versions and subsequently transformed into their discretized versions, making it feasible to tackle the IM problem as intended.
\begin{table}[t!]
\centering
\scalebox{0.85}{
\begin{tabular}{c || c| c| c| c| c} 
 \hline\hline  
 Dataset & DQSSA & DQBA & DQPSO & DBA & DPSO \\ 
 \hline
 Jazz  & 1.66 & 2.39 & 1.94 & 28.04 & 7.85 \\ 
 Wiki-Vote  & 4.91 & 5.98 & 13.37 & 70.29 & 36.54 \\
 Email-Univ & 64.32 & 23.46 & 49.87 & 185.37 & 89.48 \\
 Hamsterster & 25.88 & 12.09 & 192.76 & 517.72 & 181.12 \\[0.5ex] 
 \hline\hline
\end{tabular}}
\caption{Execution Time for baseline approaches}
\label{table:1}
\end{table}
\vspace{0pt}
\section{Experimental Setup}
\vspace{0pt}
We have evaluated our proposed method DQSSA against two kinds of baseline algorithms, centrality-based measures and MH measures, as shown in Figure \ref{QSSA}. The former includes H-index (HI), Gateway Local Rank (GLR), PageRank (PR) and Extended Neighbourhood Coreness (ENC). The latter consists of Discrete Bat Algorithm (DBA) \citep{tang2018maximizing}, Discrete Particle Swarm Algorithm (DPSO) \citep{gong2016influence}, Discrete Quantum Particle Swarm Optimisation (DQPSO), and Discrete Quantum Bat Algorithm (DQBA). While DPSO and DBA are not recent, they are still regarded as benchmarks in the field of IM. DQPSO and DQBA can be deemed up-to-date and comparable with DQSSA, since no prior research has been conducted on them in this field. These methods are evaluated on three performance metrics: Final Infected Scale (FIS), LIE Function Value and Execution Time (illustrated in Table \ref{table:1}, for the highest spreader-fraction in each dataset). The experimentation was executed on four different real-world data sets: Jazz \cite{jazz}, soc-Wiki-Vote \cite{wikivote}, email-univ \cite{email-univ} and Hamsterster. Infection probability was set to 0.1 for maximised information diffusion after extensive hyper-parameter tuning.


\section{Experimental Results and Discussion}
We observed that DQSSA surpassed all the baseline approaches in terms of performance metrics, outperforming its analogous algorithms DQBA and DQPSO, as well as the well-established algorithms DBA and DPSO, thus validating our hypothesis. It can also be observed that in terms of time complexity, DQSSA consistently outperforms majority of the state-of-the-art algorithms discussed in this study. Furthermore, DQSSA proved to be far superior in terms of influence spread as well. Therefore, this approach presents an appreciable trade-off between execution time and influence spread. Additionally, demonstrating that the fusion of quantum theory and SSA is indeed more effective than the established algorithms.
\bibliography{aaai24}
\end{document}